\documentclass[conference, onecolumn, draftclsnofoot]{IEEEtran}
\usepackage{graphicx}
\usepackage{color}
\usepackage{amsmath, multicol}
\usepackage[ruled,norelsize]{algorithm2e}
\newtheorem{theorem}{\textbf{Definition}}

\begin{document}
	\title{Matching-Game for User-Fog Assignment}
	\author{\IEEEauthorblockN{Amine Abouaomar$^{*1,2}$, Abdellatif Kobbane$^{1}$, Soumaya Cherkaoui$^{2}$
		}
		\IEEEauthorblockA{$^1$ Rabat IT Center, ENSIAS, Mohammed V University of Rabat, Morocco.\\
			$^2$ University of Sherbrooke, Canada.
			\\ 
			amine.abouaomar@ieee.org, abdellatif.kobbane@um5.ac.ma, Soumaya.Cherkaoui@usherbrooke.ca}
	}
	\maketitle
	\begin{abstract}
		Fog computing has emerged as a new paradigm in mobile network communications, aiming to equip the edge of the network with the computing and storing capabilities to deal with the huge amount of data and processing needs generated by the users’ devices and sensors. Optimizing the assignment of users to fogs is, however, still an open issue. In this paper,  we formulated the problem of users-fogs association,  as a matching game with minimum and maximum quota constraints, and proposed a Multi-Stage Differed Acceptance (MSDA) in order to balance the use of fogs resources and offer a better response time for users. Simulations results show that the performance of the proposed model compared to a baseline matching of users, achieves lowers delays for users.
	\end{abstract}
	
	\section{Introduction}
	In facing the challenges associated with huge data processing, and storage, cloud computing is now a mature technology that provides interesting features such as fault tolerance and elasticity \cite{Rachedi,Azizian1}. A new model has become possible, where resource-limited devices, especially mobile ones, can move computationally-intensive tasks to the cloud, letting device be merely used as interfaces to access services. Many research papers investigate the latency impact on sensitive applications and also the resource allocation in the cloud computing (ex. \cite{res-allo,rsu-res-mang, reconf-cloud,Azizian2}). However cloud computing has some intrinsic characteristics which make it fails to address all mobile users scenarios applications. For example, the network latency caused by far-distanced cloud servers might cause problems for delay-sensitive applications or might impair the interactive experience of the application. Also cloud computing does not offer a particular support for user mobility or take advantage of location-awareness. To cope with those shortcomings, fog computing has proposed.
	
	Fog computing has emerged as a solution to bring cloud computing features and capabilities near to the end-users, at the edge of the network. Fog computing is not a replacement for cloud computing, but a complement. Fog computing acts as cloud in terms of computing, storage and networking with two differences \cite{abouaomar1, eneya}. First, resources are located close to the user. Second, these close resources are limited compared to cloud-resources \cite{abouaomar2}. Data processing can be accelerated using fog computing but when resource-intensive tasks are needed, these can be dispatched to the cloud.
	
	With the limited resources available for Fog computing, also comes the problem of users’ assignment to fogs \cite{abouaomar2, edn}. A fog is defined as a set of resources, located close to the users, in order to answer users’ requests. Indeed, when the number of users increases, there is a tradeoff to manage between the delay experienced by users and the load among different fogs. Users’ assignment to fogs needs therefore to be investigated in order to minimize the delay experienced by users while balancing the load of fogs nodes. 
	
	In this paper we propose a solution to address the problem of users assignment for fog computing. We propose a game theoretical solution by formulating the problem as a one-to-many matching game with minimum quota constraints. The quota in our model is the number of assigned users in a fog at a given moment. In our game theoretical approach we propose a Multi-Stage Differed Acceptance (MSDA) based algorithm and we prove the classical differed acceptance (DA) algorithm is not efficient in this assignment problem due to the minimum quota rules breaking. Our algorithm uses the classical DA in several stages to reach an optimized users assignment. Our proposed algorithm allows fogs to adapt their minimum quota depending on the assigned users. This improves the network performance and optimizes the resource allocation in order to offer users a low response delay. The simulations results of the proposed solution, shows an improvement in terms of latency and load balancing.
	The paper is organized as follows:  Section II, we presents the system model, the problem formulation and the proposed algorithm; Section III is dedicated to the simulations and results.  Related works are presented in Section IV; and finally Section V ends the paper with a conclusion.

	\begin{figure}[h]
		\centering
		\includegraphics[width=.5\linewidth]{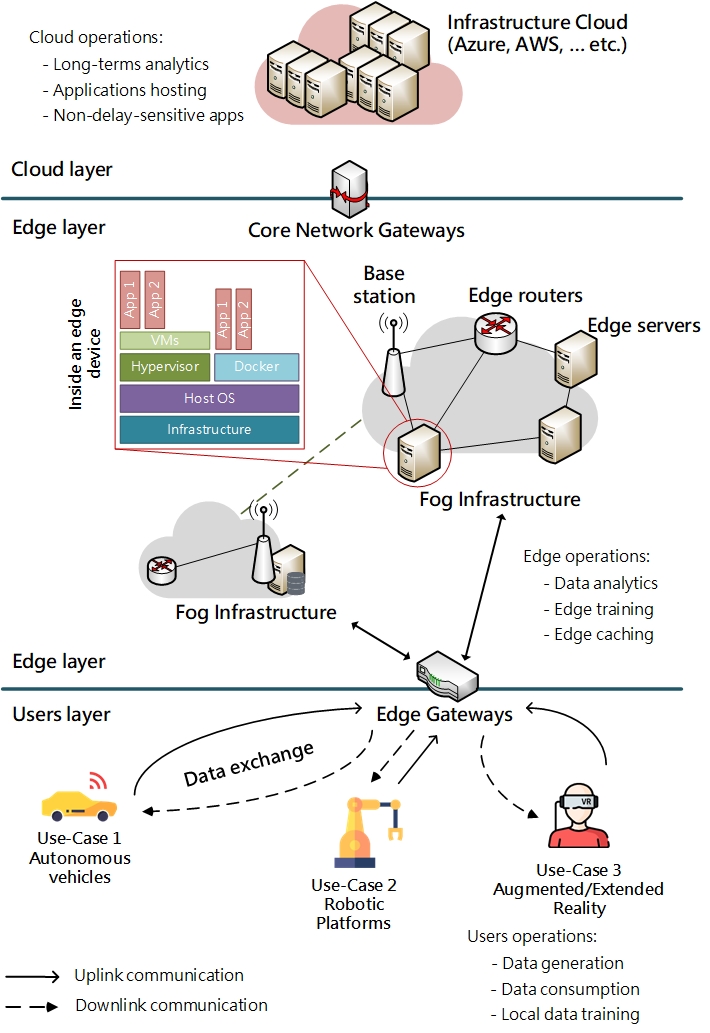}
		\caption{Architecture of a fog computing based communication network.}
		\label{fig:fog}
	\end{figure}
	\section{System Model}
	Let us consider a network where $\mathrm{U}$ users represented by the set $\mathcal{U}$. Let us consider also $\mathrm{F}$ fogs denoted $\mathcal{F}$ uniformly distributed as depicted in Fig. \ref{fig:fog}. Both users and fogs are distributed over a geographical area. Each one of the fogs has computing and storage capabilities, serving the computing tasks requests of the users. If we consider a naive users’ assignment, the users will be associated to the nearest fogs. This is not optimal in cases where a big number of users is grouped near the same fog.  Our objective in this work, is to minimize the users’ response time of the fogs while ensuring load balancing among fogs. The user response time, also called latency throughout this paper, is the delay needed for a user's request to be responded. This latency depends on the request's size, the quality of the communication channel, the size of the processing queue at the fog, and the processing capability of the fog itself. 
	
	\subsection{Latency model}
	Our primary objective is to minimize the user's response time of the fogs while ensuring their optimized resources. The latency is defined as the delay that a user's request need to be responded. It depends on the request's size, the quality of the communication channel and the size of the queue.
	Equation \ref{eq:proc_delay} describe the processing time required for each user $u$.
	\begin{equation}
		D_{u, f} = \frac{\omega_{u}}{C_f} + \tau_{queue} + \tau_{propagation}
		\label{eq:proc_delay}
	\end{equation}
	where $\omega_{u}$ describes the user's $u$ task size, and $C_{f}$ denotes the maximum processing performance of the fog and $\tau_{queue}$ is the waiting delay in the queue. The propagation delay $\tau_{propagation}$ is the sum of the delay needed for a request to reach the fog and the delay needed for a response to reach the user. $\tau_{propagation}$ depends on the network state and the bandwidth. It is described by the following formula:
	\begin{equation}
		\tau_{propagation} = bw_{u, f} . \log_2(1 + SINR_u)
		\label{eq:propg}
	\end{equation}
	Where $bw_{u, f}$ denotes the bandwidth and $SINR_u$ is the quality of channel between the user and the fog.
	The queuing time is the previous requests processing time, represented in the following equation:
	\begin{equation}
		\tau_{queue} = \sum_{i \in \mathcal{U}}\frac{\omega_{i}}{C_f}
		\label{eq:queuing}
	\end{equation}
	Where $\omega_{i}$ is the $i^{th}$ user's request task size, and $C_f$ is the fog's maximum processing performance.  
	\subsection{Problem formulation}
	The problem of users assignment is considered as a decision making policy that we denote $\mu$, that for each pair $(u, f)$, $u$ in $\mathcal{U}$ and $f$ in $\mathcal{F}$, returns a binary value $x_{u,f}$ in $\{0, 1\}$, that indicates either the user $u$ is associated to the fog $f$ or not.
	
	\begin{gather}
		x_{u, f} = 
		\begin{cases}
			x_{u, f} = 1~:\text{if the user $u$ is assigned to fog $f$}\\
			x_{u, f} = 0~:\text{if not}
		\end{cases}
		\label{eq:policy}
	\end{gather}
	The fog load is defined as follow :
	\begin{equation}
		p_{f} = \sum_{u \in \mathcal{U}} x_{u, f} . D_{u,f}
		\label{eq:fog_load}
	\end{equation}
	The quality of the load balancing depends on the maximum load difference of the fog, which is defined as the difference between the maximum and the minimum load of the fog.
	From the equation (\ref{eq:fog_load}) we define $\delta_{p}$ the maximum load difference of a fog using a policy $\mu$ as follow :
	\begin{equation}
		\delta_{p}(\mu) = max(p_{f}) - min(p_{f})
		\label{eq:diff_fog_load}
	\end{equation}
	a better load balancing requires a smaller $\delta_{p}$. The problem can be formulated by:
	\begin{subequations}
		\begin{equation}
			minimize : \sum_{\mathrm{f} \in \mathcal{F}} \sum_{\mathrm{u} \in \mathcal{U}} x_{u, f} D_{u, f}
			\label{eq:load}
		\end{equation}
		\begin{equation}
			\sum_{\mathrm{f} \in \mathcal{F}} x_{u, f} \leq 1,
			\label{eq:assoc}
		\end{equation}
		\begin{equation}
			\text{with : } x_{u, f} \in \{0, 1\},
			\label{eq:assoc-index}
		\end{equation}
		\begin{equation}
			\mathrm{p_f} \ge \mathrm{q_{f}^{min}}, \quad \forall \mathrm{f} \in \mathcal{F},
			\label{eq:quota_min}
		\end{equation}
		\begin{equation}
			\mathrm{p_f} \le \mathrm{q_{f}^{max}}, \quad \forall \mathrm{f} \in \mathcal{F}
			\label{eq:quota_max}
		\end{equation}
	\end{subequations}
	The $q_{f}^{max}$ and $q_{f}^{min}$ denotes the minimum and the maximum quota for a fog $f$. In other words, it is the maximum and the minimum number of users the fog $f$ can serve.
	
	\subsection{Game formulation}
	Before presenting our game theory-based solution, it is important to highlight the fact that the above stated problem is of exponential complexity $(O(F^U))$ in case we implement an exhaustive search of solutions. Since we are in a system with several users and each user has their own requirements, a system with exponential complexity is not feasible as it would consume considerable resources, leading to a high latency.
	
	Game theory offers a decentralized powerful mathematical tool with low complexity for combinatorial problems (ex., (\ref{eq:load}) and (\ref{eq:assoc-index})).
	The problem is formulated as a matching game, that takes in account the minimum quota and the maximum quota constraints presented in (\ref{eq:quota_min}) and (\ref{eq:quota_max}).	
	
	A matching game, from a general point of view, is a two-sided assignment problem between two disjoint sets of players, and each one of the players of each set is interested to be matched to the players of the other set. In our scenario, the two sets of players of the matching game are $\mathcal{U}$ and $\mathcal{F}$. A preference relation $\succ$, which is binary, is defined between the elements of a given set. Note that this relation is complete, reflexive, and transitive.
	
	We denote $\succ_{u}$ the preference relation of the user $u \in \mathcal{U}$. $\mathrm{f} \succ_{u} \mathrm{f'}$ means that the user $u$ prefer the fog $f$ over the fog $f'$, $f$ and $f' \in \mathcal{F}$. Using the same notation, we define the preferences of the fogs over the users by $\succ_{f}$.
	
\begin{figure}
		\centering
		\includegraphics[width=.7\linewidth]{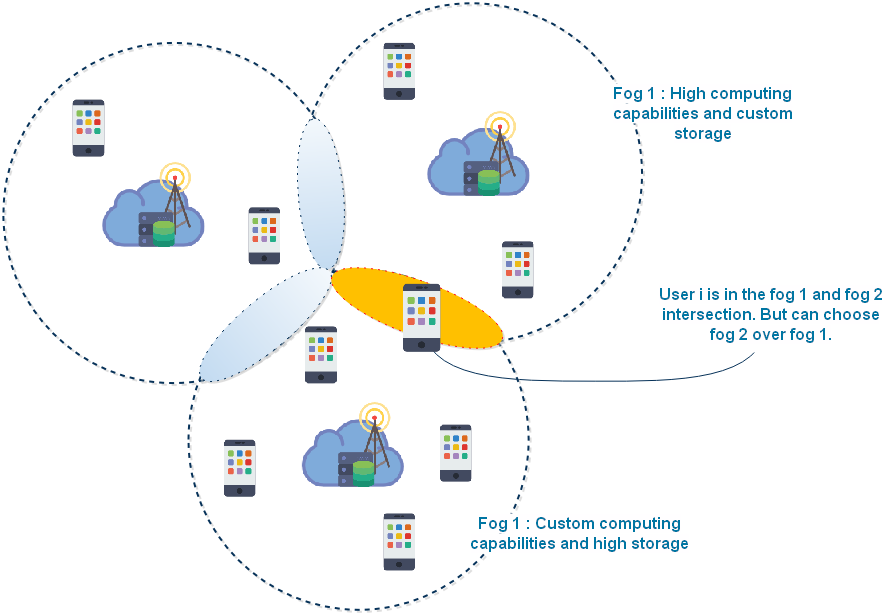}
		\caption{Users distribution over a geographical area, covered by three fogs networks. The user in the intersection zone between the fog 1 and the fog 2, can choose to be associated to fog 2 or fog 1, depending on his needs (storage or computing in this case)}
		\label{fig:inter}
	\end{figure}

	Note here, that when adopting a preferences based system, the criteria used for preferences can change from user to another. Fig. \ref{fig:inter} Shows a use case, when a number of users are distributed over an area covered by a number of fogs networks. For the example shown in Fig. \ref{fig:inter}, and in a naive case, where there is no consideration to preferences, the user in the intersection zone between the fogs (orange zone) will be associated to the nearest fog, because the signal power is high. But in case where the user is not interested in computing but more in storage, they can choose to be connected to the other fog offering storage capabilities, even of the signal is not that high.

	\subsection{Matching game with minimum quota}
	The preference relations are defined using utility functions of each entity of the system. For the users, the utility function depends on the latency, while for the fogs the utility function depends on optimal resources utilization. Matching game based associations provides a good way to optimize resource usage by adjusting the min/max quota. For our problem, we define a one-many matching game as follows:
	\begin{theorem}[One-to-many matching game]\\
		A one-to-many matching $\mu$ is defined as a mapping from the set of the users $\mathcal{U}$ to $\mathcal{F}$, that satisfy :
		\begin{itemize}
			\item $\forall u \in \mathcal{U}$, \quad $\mu(u) \subseteq \mathcal{F}$
			\item $\forall f \in \mathcal{F}$, \quad $\mu(u) \in \mathcal{U}$
			\item $\mu(f) = u$ exists if and only if $u \in \mu(f)$
		\end{itemize}
	\end{theorem}
	If we consider the case where $\mu(f) = u$, the user can be assigned or not, which means that, either $x_{u, f}=1$ or $x_{u, f}=0$. The matching satisfies the constraints in (\ref{eq:assoc-index}) and (\ref{eq:assoc}). The matching $\mu$ is considered to be feasible if it satisfy the quota constraints $p_f^{min}$ $\le$ $p_f$ $\le$ $p_f^{max}$, with $p_f = |\mu(f)|$ and of course $|\mu(u)| \in \{0, 1\}$.
	To proove the stability of the matching game, we demonstrate that there is no blocking pair. A blocking pair is defined as follows :
	\begin{theorem}[Blocking pair]\\
		A pair ($u$, $f$), $u \in \mathcal{U}$ and $f \in \mathcal{F}$ is blocking pair, if and only if :
		\begin{itemize}
			\item $f$ $\succ_{u} \mu(u)$
			\item $u \succ_{f} u'$ for some $f' \in \mu(u)$
		\end{itemize}
	\end{theorem}
	
	Taking in consideration the satisfaction of the quota defined in (\ref{eq:quota_min}) and (\ref{eq:quota_max}) the matching is feasible. The fairness is guaranteed, since no pair ($u$, $f$), $u \in \mathcal{U}$ and $f \in \mathcal{F}$ is blocking pair.
	
	\section{Simulation results}
	To test our model we considered the following parameters :
	\begin{itemize}
		\item $|\mathcal{F}| = 5 $ fogs (Having different processing capabilities).
		\item $|\mathcal{U}| = 500$ users requesting between $5$ to $15$ tasks.
		\item The minimum quotas are chosen between $0$ and $\frac{|N|}{|F|}$.
	\end{itemize}
	We evaluate the systems and compare it performances with a random users’ assignment policy with no consideration to the quota constraints (called baseline scheme) . We evaluate : 
	\begin{itemize}
		\item The impact of number of users on the delay.
		\item The load of each fog while the system is using a simple baseline scheme for users assignment and the proposed model based on the multi-staged differed acceptance algorithm with simple.
		\item The delay evolution in function of the number of users.
		\item The load impact on the delay.
	\end{itemize}
	\subsection{Proposed algorithm}
	The classical differed acceptance algorithm (DA) does not support the minimum quota parameter. Let us consider a system with the set of users $\mathcal{U}=\{u_1, u_2, u_3\}$, and a set of fogs $\mathcal{F}=\{f_1, f_2, f_3\}$ and assuming that the minimum quota ($q_f^{min}$) is $1$ and the maximum quota ($q_f^{max}$) is $2$ for all the fogs. Considering a general preferences relation $h$. If we consider that $u_1 \succ_{h} u_2 \succ_{h} u_3$ and $f_1 \succ_{u_i} f_2 \succ_{u_i} f_3$ for all $u_i$ $\in$ $\mathrm{U}$,  applying the classical DA means that (1) $\mu(f_1) = {u_1, u_2}$; (2) $\mu(f_2) = {u_3}$; (3) $\mu(f_3) = \emptyset$. This break the minimum quota rules.
	
	The proposed algorithm is inspired from \cite{strategyproof}. The multistage deferred acceptance algorithm (MSDA) is an algorithm that is run in several stages or steps. We assume that all the fogs share a list of preference, that rank all the users following . We call it this list the global preference list $GL$, and we denote $\succ_{gl}$ the preference relation. In the initial stage, we temporarily reserve a subgroup of users and turn on the classical DA on the remaining subgroup.  The assignments from the given stage are made final, and we reduce the minimum and maximum quotas accordingly.
	
	In every stage we classify the user into the most preferred and the less preferred. Note that the number of the less preferred users is the sum of the minimum quota of all the fogs. Depending on the number of preferred and non-preferred, users we choose to run the classical differed acceptance algorithm (DA) with maximum quota. If the set of the most wanted users is empty we run the DA with the minimum quota values, otherwise we run it with the maximum quota.
	\begin{algorithm}
		\caption{Users assignment with min/max quota}
		\label{algo:quota-min-algo}
		\SetKwInOut{Input}{Input}
		\SetKwInOut{Output}{Output}
		\Input{$\succ_{GL}, \succ_f, \succ_u, q_{f}^{min}, q_{f}^{max}$}
		\Output{$\mu$ : the global users assignment policy}
		\textbf{Initialization : }\\
		- $\mu = \emptyset$, $r^0 = GL$, $q_{f}^{1,min} = q_{f}^{min}$, $q_{f}^{1,max} = q_{f}^{max}; (\forall f \in  \mathcal{F})$\\
		\textbf{Policy construction :}\\
		\Repeat{$GL = \emptyset$}{
			$r^k$ = $\sum_{i=1}^{|q_{f}^{min}|} q_{f_i}^{min}$\\
			$R^k$ = ${s_i}$ where $s_i$ is a less wanted user following $GL$ ranking. \\
			\eIf{$R^{k-1} \backslash R^k \ne \emptyset$}{
				- invoke the classic DA on $R^{k-1} \backslash R^k$ with the maximum quota = $q^{k,max}_f$\\
				- revoke the matched users from $GL$\\
				- add the matched users to $\mu$.
			}{
				- invoke the classic DA on $R^k$ with the maximum quota = $q^{k,min}_f$\\
				- revoke the matched users from $GL$\\
				- add the matched users to $\mu$.
			}
			- update the maximum quota : $ q_{k,max}^f = q_{k-1,min}^f - \sum_{i=1}^{|\mu_k|} D_{u_i,f}$ where $\mu_k$ is the matching at the stage $k$ \\
			- update the minimum quota : $q_{k,min}^f = max\{0, q_{k-1,min}^f - \sum_{i=1}^{|\mu_k|} D_{u_i,f}\}$ \\
		}
	\end{algorithm}
	$R^{k-1} \ R^k$ denotes the difference between the set of the well ranked users from two successive stages.
	
	\begin{figure}[h]
		\centering
		\includegraphics[width=.7\linewidth]{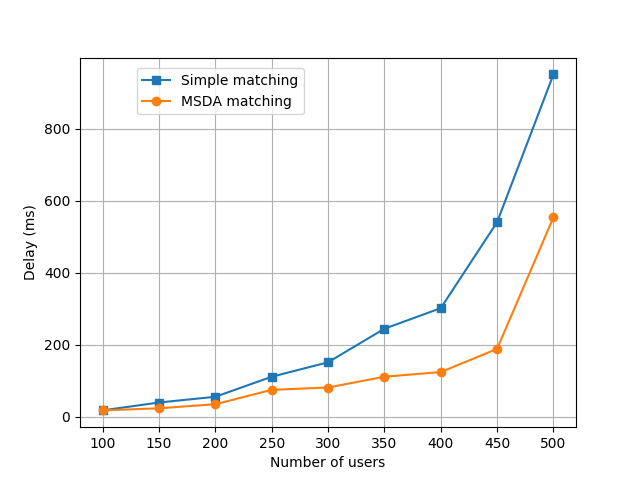}
		\caption{The delay evolution in the proposed scheme versus the baseline model. The delay stay low even with a high number of users.}
		\label{fig:figure1}
	\end{figure}
	\begin{figure}[h]
		\centering
		\includegraphics[width=.7\linewidth]{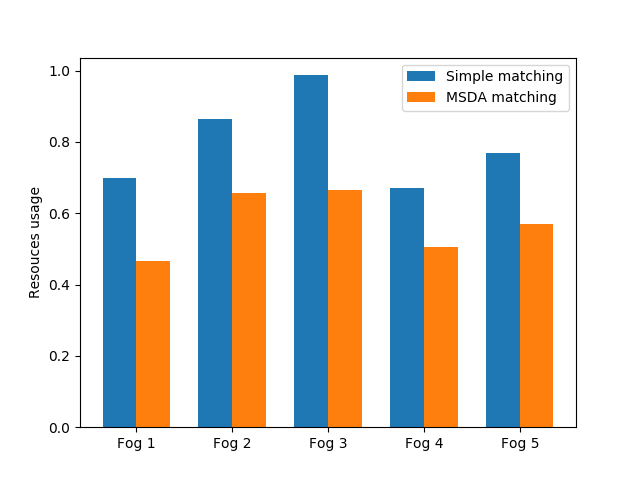}
		\caption{Comparison of resource usage of each fog in the baseline scheme and the proposed approach, in case both the systems are processing the same requests.}
		\label{fig:figure2}
	\end{figure}
	\begin{figure}[h]
		\centering
		\includegraphics[width=.7\linewidth]{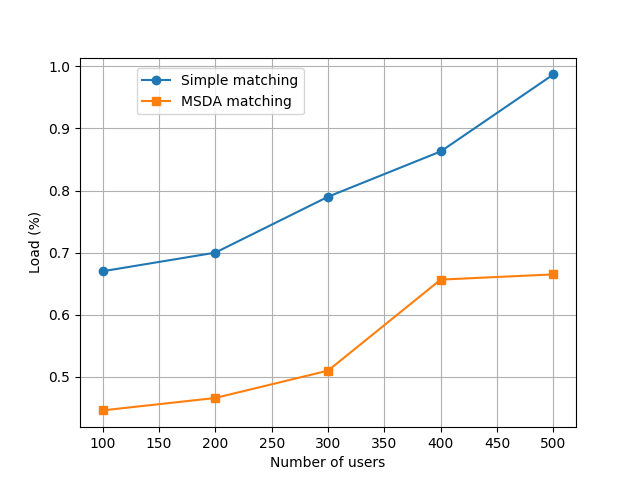}
		\caption{Load of fogs' resources }
		\label{fig:figure3}
	\end{figure}
	
	From a first view to Fig. \ref{fig:figure1}, we see that the proposed model achieves a good improvement in term of delay, even with a high number of users. This is because the association is based on taking in account the preferences of the entities of the system, instead of adopting random assignment policies based on criteria like the distance and the signal strength. The proposed approach takes as input the preferences and based on these, and the state of the system (i.e. number of users, minimum and maximum quota, etc.) it try to find an optimal assignment policy to have a better load balancing.
	
	Fig. \ref{fig:figure1} illustrates the variation of the delay for he proposed algorithm based on multi-stage differed acceptance and the baseline scheme It is shown that the proposed algorithm is more efficient and guarantees a low latency (up to 30\% in some cases), offering the users a better response. The preferences based systems allows the users to rank the fogs, and also the fogs to rank the users. In this way, users can choose a fog even if it is not the closest one. 
	
	Fig. \ref{fig:figure2} shows the resources usage of the different fogs while processing the same number of requests. Since the baseline scheme does not take into consideration any load balancing among fogs, crowded fogs suffer congestion, leading to a high latency for users. The proposed algorithm guarantees load balancing by taking in account the quota parameters (minimum and maximum quota) giving the system the ability to have a better load balancing and a better users’ assignment. The fact that the assignment policy changes depending on the ranking of users and the quota, guarantee for the users a better response time and a good resource utilization for the fogs.
	
	Fig. \ref{fig:figure3} explains the fogs' resources utilization when varying the number of users. In the baseline scheme, the load increases quickly especially with a high number of assigned users, affecting the delay in a negative way. 
	From the previous simulations, we can see that when there is a bad load balancing (random users’ assignment scheme) the delay increases, which may even discourage the users from using the services. On the contrary, when we adopt a preferences -based matching approach which takes into account the quota constraints, the response time decreases for users.  Therefore, users benefit from services with low latency and the fogs resources usage is more efficient.
	
	\section{Related Works}
	The problem of user assignment was widely investigated in various papers (ex. \cite{assoc-hetnets} and \cite{load-balance}), but few works investigated it from the fog computing perspective.
	
	Chakroun et al. in \cite{reconf-cloud} proposed a new algorithms to optimize Mobile Cloud Computing resources management techniques based on a stochastic networks optimization. The proposed algorithm take in account the energy consumption on the cloud data center side while ensuring resources elasticity to adapt to users’ demands and insure a highly available platform. And in \cite{res-allo}, they proposed an approach for reducing users’ requests delay while enhancing resource availability in Mobile Cloud Computing resources allocation using stochastic networks optimization based on Lyapunov optimization.
	
	Liu et al. in \cite{user-assoc-survey} proposed a taxonomy and an extensive overview of the state-of-the-art in user association algorithms conceived for HetNets, massive MIMO, mmWave, and energy harvesting networks. Indeed, to make the mobile networks supports the huge data generated by the devices (users, IoT, ...), key enabling technologies, such as heterogeneous networks (HetNets), massive multiple-input multiple-output (MIMO), and millimeter wave (mmWave) techniques, have been identified to bring 5G to fruition. User association plays the cornerstone role in optimizing the load balancing, the spectrum efficiency, and the energy efficiency of networks. Since research efforts are dedicated to the problems of user association in HetNets, massive MIMO networks, mmWave networks, and energy harvesting networks.
	
	Semiari et al. in \cite{load-balance} proposed a new cell association solution, that considers the achievable rate to assign users to $mmW-BSs$ or $\mu W-BSs$. They proposed a one-to-many matching game based solution that takes in consideration the minimum quota constraints for the BSs, their model shows an efficient way to balance the load over the $mmW$ and $\mu W$ frequency bands.
	
	Cacciapuoti in \cite{mobility-aware-assoc} proposed a mobility-aware user association strategy for mmW networks to face the limitations of the conventional received power (RSS)-based association strategies. The proposed strategy is fully distributed takes in consideration the load distribution (i.e : overcoming the association of users to an already congested small base station). Also they designed a polynomial-time complexity algorithm to face the problem of exhaustive search for the optimal solution for the association problem.
	
	Zhao et al. proposed in \cite{user-ap} a predictive scheduling in wireless caching network for the user-AP association and dynamic resource allocation problem \cite{Azizian2}. by exploiting the structure information of the problem they proposed an online algorithm called Predictive User-AP association and Resource Allocation it guarantee the performance and does not require any statistical  information of the system dynamics.
	
	Tung et al. in \cite{fran} proposed a scheme to joint the user association, data delivery rate and signal precoding in the downlink of a cache-enabled F-RAN with limited fronthaul capacity. They formulated an optimization problem in order to maximize the weighted difference of network throughput and total power consumption.
	\section{Conclusion}
	In this paper we tackled the problem of users’ assignment for fog computing. We proposed a matching game algorithm with minimum and maximum quota constraints between the users and the fogs, with the goal of balancing fogs resources usage and offering a low response delay for users. Simulations results show that the performance of the proposed model compared to a baseline matching and a random assignment policy offers lower delays for users and better balancing of resource use at fogs.
	

\begin{thebibliography}{9}
		
		\bibitem{res-allo}
		O. Chakroun et al., "\emph{Resource Allocation for Delay Sensitive Applications in Mobile Cloud Computing,}" 2016 IEEE 41st Conference on Local Computer Networks (LCN), Dubai, 2016, pp. 615-618.
		
		\bibitem{rsu-res-mang}
		M. A. Salahuddin et al., "\emph{RSU cloud and its resource management in support of enhanced vehicular applications}", 2014 IEEE Globecom Workshops (GC Wkshps), Austin, TX, 2014, pp. 127-132.
		
		\bibitem{reconf-cloud}
		O. Chakroun et al., "\emph{Reducing Energy Consumption for Reconfiguration in Cloud Data Centers}", 2016 IEEE 84th Vehicular Technology Conference (VTC-Fall), Montreal, QC, 2016, pp. 1-6.
		
		\bibitem{strategyproof}
		D. Fragiadakis et al., "\emph{Strategyproof Matching with Minimum Quotas}".
		
		\bibitem{user-assoc-survey}
		D. Liu et al., "\emph{User Association in 5G Networks: A Survey and an Outlook}", in IEEE Communications Surveys \& Tutorials, vol. 18, no. 2, pp. 1018-1044, Secondquarter 2016.
		
		\bibitem{load-balance}
		O. Semiari et al., "\emph{Downlink Cell Association and Load Balancing for Joint Millimeter Wave-Microwave Cellular Networks}", 2016 IEEE Global Communications Conference (GLOBECOM), Washington, DC, 2016, pp. 1-6.
		
		\bibitem{abouaomar2}
		A. Abouaomar et al., "\emph{Caching, device-to-device and fog computing in 5th cellular networks generation : Survey,}" 2017 International Conference on Wireless Networks and Mobile Communications (WINCOM), 2017, pp. 1-6, doi: 10.1109/WINCOM.2017.8238174.
		
		\bibitem{mobility-aware-assoc}
		A. S. Cacciapuoti et al., "\emph{Mobility-Aware User Association for 5G mmWave Networks}", in IEEE Access, vol. 5, pp. 21497-21507, 2017.
		
		\bibitem{user-ap}
		S. Zhao et al., "\emph{Online User-AP Association with Predictive Scheduling in Wireless Caching Networks}", IEEE Global Communications Conference (GLOBECOM) 2017
		
		\bibitem{fran}
		T. T. Vu et al., "\emph{Joint Optimization of User Association, Data Delivery Rate and Precoding for Cache-Enabled F-RANs}", IEEE Global Communications Conference (GLOBECOM) 2017.
		
		\bibitem{assoc-hetnets}
		A. Benmimoune et al., "\emph{User Association for HetNet Small Cell Networks}", 2015 3rd International Conference on Future Internet of Things and Cloud, Rome, 2015, pp. 113-117.
		
		\bibitem{Azizian2}
		M. Azizian et al., "\emph{An Optimized Flow Allocation in Vehicular Cloud,}" in IEEE Access, vol. 4, pp. 6766-6779, 2016, doi: 10.1109/ACCESS.2016.2615323.
		
		\bibitem{abouaomar1}
		A. Abouaomar et al., "\emph{Users-Fogs association within a cache context in 5G networks:Coalition game model,}" 2018 IEEE Symposium on Computers and Communications (ISCC), 2018, pp. 00014-00019, doi: 10.1109/ISCC.2018.8538500.
		
		\bibitem{eneya}
		Z. Ennya et al., "\emph{Computing Tasks Distribution in Fog Computing: Coalition Game Model,}" 2018 6th International Conference on Wireless Networks and Mobile Communications (WINCOM), 2018, pp. 1-4, doi: 10.1109/WINCOM.2018.8629587.
		
		\bibitem{edn}
		E. D. N. Ndih et al., "\emph{On Enhancing Technology Coexistence in the IoT Era: ZigBee and 802.11 Case,}" in IEEE Access, vol. 4, pp. 1835-1844, 2016, doi: 10.1109/ACCESS.2016.2553150.
		
		\bibitem{Azizian1}
		M. Azizian et al., "\emph{Vehicle Software Updates Distribution with SDN and Cloud Computing,}" in IEEE Communications Magazine, vol. 55, no. 8, pp. 74-79, Aug. 2017, doi: 10.1109/MCOM.2017.1601161.
		
		\bibitem{Rachedi}
		A. Rachedi et al., "\emph{IEEE Access Special Section Editorial: The Plethora of Research in Internet of Things (IoT),}" in IEEE Access, vol. 4, pp. 9575-9579, 2016, doi: 10.1109/ACCESS.2016.2647499.
		
		
	\end{thebibliography}
\end{document}